

\input phyzzx

\def\e{\epsilon}

\def\l{\lambda}

\def\f{\varphi}

\def\P{\Psi}

\def\S{\Sigma}

 \def\noss{\noalign{\smallskip}}
\def\noms{\noalign{\medskip}}

\def\dal{\sqcup\kern-.29cm{\sqcap}}
\def\lap{\buildrel \leftarrow\over\partial}
\def\rap{\buildrel \rightarrow\over\partial}

\def\ft#1#2{{\textstyle{{#1}\over{#2}}}}
\def\frac#1#2{{{#1}\over{#2}}}
\def\1#1{\frac1{#1}} \def\2#1{\frac2{#1}} \def\3#1{\frac3{#1}}
\def\noss{\noalign{\smallskip}}
\def\sb#1{\lower.4ex\hbox{${}_{#1}$}}

\def\pa{\partial}

\def\.{\,\,,\,\,}
\def\'{\mkern 1mu}


\def\crampest{\medmuskip = 1mu plus 1mu minus 1mu}
\def\uncramp{\medmuskip = 4mu plus 2mu minus 4mu}

\def\IR{{\hbox{{\rm I}\kern-.2em\hbox{\rm R}}}}
\def\IB{{\hbox{{\rm I}\kern-.2em\hbox{\rm B}}}}
\def\IN{{\hbox{{\rm I}\kern-.2em\hbox{\rm N}}}}
\def\IC{{\ \hbox{{\rm I}\kern-.6em\hbox{\bf C}}}}

\def\IZ{{\hbox{{\rm Z}\kern-.4em\hbox{\rm Z}}}}
\def\to{\rightarrow}
\def\underarrow#1{\vbox{\ialign{##rcr$\hfil\displaystyle
{#1}\hfil$rcr\noalign{\kern1pt
\nointerlineskip}$\longrightarrow$rcr}}}
%

\tolerance=5000
\overfullrule=0pt

\twelvepoint
\rightline{UG-8/92}
\rightline{CERN-TH.6647/92}
\rightline{UCB-PTH-92/27}
\rightline{LBL-32806}
\date{\ }
\titlepage
\title{A Derivation of the BRST Operator for Non-Critical
$W$ Strings}
\vglue-.25in
\author{E. Bergshoeff\foot{Institute for Theoretical Physics,
University
of Groningen, Nijenborgh 4, 9747 AG Groningen, The Netherlands.},
A. Sevrin\foot{Theoretical Physics Group, Lawrence Berkeley
Laboratory,
1 Cyclotron Rd. and Department of Physics, University of California
at
Berkeley, Berkeley, CA 94720, USA.}  and X. Shen\foot{Theory
Division, CERN, CH-1211
Geneva 23, Switzerland.\hfill\break
\medskip\noindent{\twelvepoint CERN-TH.6647/92\hfill\break
September 1992\hfill\break}}}

\bigskip
\abstract{We derive the recently proposed BRST charge for
non-critical $W$ strings from a
Lagrangian approach. The basic observation is that, despite
appearances, the
combination of two classical ``matter'' and ``Toda''
$w_3$ systems leads to a closed
modified gauge algebra, which is of the so-called
soft type.
Based on these observations, a novel way to construct critical
$W_3$ strings is given.}

\vskip 1.5in
\centerline{\it Dedicated to Professor F. Cerulus on the Occasion of
his 65th
Birthday}

\endpage

\REF\wstring{A. Bilal and J.-L. Gervais,
Nucl. Phys. {\bf B326} (1989) 222;
S. Das, A. Dhar and S.~ Rama, Int. J. Mod. Phys. {\bf A7} (1992) 2295.}
\REF\po{C.N. Pope, L.J. Romans and K.S. Stelle, Phys. Lett.
{\bf 269B} (1991) 287;
C.N. Pope, L.J. Romans, E. Sezgin and K.S. Stelle,
Phys. Lett. {\bf B274}
(1992) 298; H. Lu, C.N.~ Pope, S. Schrans and K.W. Xu, preprint CTP
TAMU-5/92, KUL-TF-92/1;
H. Lu, C.N. Pope, S. Schrans and X.J. Wang, preprint
CTP-TAMU-10/92, KUL-TF-92/8 and
Mod. Phys. Lett. {\bf A7} (1992) 1835.}
\REF\za{A.B. Zamolodchikov, Theor. Math. Phys. {\bf 65} (1986) 1205.}
\REF\tmieg{J. Thiery-Mieg, Phys. Lett. {\bf B197} (1987) 368.}
\REF\ssvann{K. Schoutens, A. Sevrin and P. van Nieuwenhuizen, Commun.
Math. Phys. {\bf 124} (1989)~ 87.}
\REF\hull{C.M. Hull, Phys. Lett. {\bf B240} (1990) 110.}
\REF\fz{V.A. Fateev and A.B. Zamolodchikov,
Nucl. Phys. {\bf B280} [FS18]
(1987) 644.}
\REF\romans{L.J. Romans, Nucl. Phys. {\bf B352} (1991) 829.}
\REF\blnw{M. Bershadsky, W. Lerche, D. Nemeschansky and N.P. Warner,
preprint
CERN-TH.6582/92.}
\REF\ossvan{H. Ooguri, K. Schoutens, A. Sevrin and P. van
Nieuwenhuizen, Commun. Math. Phys. {\bf 145} (1992) 515.}
\REF\ssvan{K. Schoutens, A. Sevrin and P. van Nieuwenhuizen, Nucl.
Phys. {\bf B371} (1992) 315.}
\REF\ssvannn{K. Schoutens, A. Sevrin and P. van Nieuwenhuizen, Phys.
Lett. {\bf B243} (1990) 245.}
\REF\bv{J.A. Batalin and G.A. Vilkovisky, Phys. Rev. {\bf D28} (1983)
2567 and
{\bf D30} (1984) 508; Nucl. Phys. {\bf B234} (1984) 106.}
\REF\proeyen{A. van Proeyen, preprint-KUL-TF-91/35, to appear in the
proceedings of ``Strings and Symmetries 1991;''  W. Troost,
P. van Nieuwenhuizen and A.~ van Proeyen, Nucl. Phys. {\bf B333}
(1990) 727;
W. Troost and A. van Proeyen, book in preparation.}
\REF\twelve{E.S. Fradkin and M.A. Vasiliev, Phys. Lett. {\bf B72}
(1977)70;
G. Sterman, P.K. Townsend and P. van Nieuwenhuizen, Phys. Rev. {\bf
D17} (1978)
1501; R.G. Kallosh,  v. Pis'ma Zh. Eksp. Teor. Fiz. {\bf 26} (1977) 575;
Nucl.
Phys. {\bf B141} (1978) 141; B. de Wit and J.W.~ van Holten, Phys.
Lett. {\bf
B79} (1978) 389; B. de Wit, P. van Nieuwenhuizen and A. van~ Proeyen,
Phys.
Lett. {\bf B104} (1981) 27.}
\REF\hhull{C.M. Hull, Nucl. Phys. {\bf B367} (1991) 731.}
\REF\ppoo{C.N. Pope, L.J. Romans and K.S. Stelle, Phys. Lett. {\bf 268B}
(1991) 167.}
\REF\poly{A.M. Polyakov, Mod. Phys. Lett. {\bf A2} (1989) 893.}
\REF\sev{K. Schoutens, A. Sevrin and P. van
Nieuwenhuizen, preprint ITP-SB-91-13, to appear in the proceedings of
the
Workshop on Quantum Field Theory, Statistical Mechanics,
Quantum Groups and Topology, Coral Gables, 1991.}
\REF\minustwo{P. Bouwknegt, A. Ceresole, P. van Nieuwenhuizen and J.
McCarthy, Phys
Rev. {\bf D40} (1989) 415.}
\REF\thiel{K. Thielemans, Int. J. Mod. Phys. {\bf C2} (1991) 787.}
\REF\bss{E. Bergshoeff, A. Sevrin and X. Shen, in preparation.}

\thispage1
\smallskip
\noindent
{\bf 1.\ Introduction}
\smallskip

So far, all attempts to construct {\it critical}
$W_3$ strings [\wstring, \po]
have concentrated on the standard $W_3$ algebra of Zamolodchikov
[\za]. The corresponding BRST charge, which plays an essential role
in the determination of the physical state spectrum, was constructed
some time ago by Thiery-Mieg [\tmieg]. This result has been generalized
in [\ssvann] to arbitrary quadratically generated non-linear algebras,
while in [\hull] the BRST charge was derived from a Lagrangian
point of view. The construction of the BRST charge involves the
introduction of ghosts and antighosts for the Virasoro and the
spin-3 symmetries.
It turns out that nilpotency of the BRST charge
requires that the matter contribution $c_M$ to the central charge
be $c_M=100$. One therefore has to search for matter realizations
of the $W_3$ symmetry at this value of the central charge.
The standard realization
of Fateev and Zamolodchikov [\fz] involves only two free scalars,
but more would be preferred in view of string applications. In [\romans],
realizations involving an arbitrary number of scalars were given. The
application of this multi-scalar realization in the context of critical
$W_3$ strings was discussed in [\po].

\smallskip
Recently, Bershadsky, Lerche, Nemeschansky and Warner [\blnw] constructed
a BRST charge for a {\it non-critical} $W_3$ string, i.e. for a $W_3$
minimal model coupled to $W_3$ gravity. This BRST charge differs from
the one of [\tmieg]. The construction of [\blnw] is an important
step forwards in the study of non-critical $W_3$ strings, which are
expected to be exactly solvable. An earlier
investigation of non critical $W_3$ strings was given in
[\ossvan, \ssvan], where both the induced
and the
effective $W_3$-gravity Lagrangians were obtained and the $W_3$ KPZ
formula was
derived.
In the analysis of [\blnw] one starts with
two mutually commuting $W_3$ algebras generated by the
energy-momentum tensors
$T^{(i)}$ and the spin-3 currents $W^{(i)}\  (i=1,2)$.
The $i=1$ copy represents the matter sector while the $i=2$ copy
represents the $W_3$-Liouville (Toda) sector, which does not decouple
in the non-critical case.
Both satisfy the quantum
$W_3$ algebra with operator product expansions (OPE) given by
$$
\eqalign{
T^{(i)}(z) T^{(i)}(w) &\sim {c_i/2\over (z-w)^4} + {2T^{(i)}\over
(z-w)^2}
+{\pa T^{(i)}\over (z-w)},\cr
\noms
T^{(i)}(z) W^{(i)}(w) &\sim {3W^{(i)}\over (z-w)^2} +{\pa
W^{(i)}\over
(z-w)}, \cr}
$$
\smallskip
$$
\eqalign{
W^{(i)}(z) W^{(i)}(w) \sim &{c_i\over 9\beta_i^2(z-w)^6}+
{2T^{(i)}\over 3\beta_i^2 (z-w)^4} +{\pa T^{(i)}\over 3\beta_i^2
(z-w)^3}\cr
&\ +{1\over (z-w)^2}\Big(\ft23\Lambda^{(i)} + \ft1{10\beta_i^2}\pa^2
T^{(i)}\Big)
+{1\over (z-w)}\Big(\ft13\pa\Lambda^{(i)}+\ft1{45\beta_i^2}\pa^3
T^{(i)}\Big),\cr}\eqn\wthree
$$
where
$$
\beta_i^2\equiv {16\over 5c_i+22},\eqn\two
$$
and
$$
\Lambda^{(i)}=(T^{(i)} T^{(i)})-\ft3{10}\pa^2 T^{(i)}.\eqn\three
$$
For later purposes, it is convenient to use
a non-standard normalization of the spin-3 generators such that
the coefficient in front of the composite $\Lambda$-term above is
independent of the central charge.

The sum of the matter and Liouville sytem, i.e.
$T^{\rm tot}=T^{(1)}+T^{(2)}$ and
$W^{\rm tot}=W^{(1)}+W^{(2)}$, does not form a closed $W_3$ algebra.
One may verify
that there is no way to construct a $W_3$ algebra out of
$T^{(i)}$ and $W^{(i)}$ only. The remarkable result of [\blnw] is that
nevertheless it is possible to construct a nilpotent BRST charge for the
matter and Liouville sytems.
Introduce ghosts and antighosts $c^{(1)}, b^{(1)},c^{(2)}, b^{(2)}$
for
the Virasoro and the spin-3 symmetry, respectively. The BRST charge
of [\blnw] is then given by
$$
Q={1\over 2\pi i}\oint \big\lbrace c^{(1)}\big(T^{(1)}+T^{(2)}+\ft12
T_{\rm gh}\big) + c^{(2)}\big(W^{(1)}\pm iW^{(2)}+\ft12
W_{\rm gh}\big)\big\rbrace,\eqn\brst
$$
where $T_{\rm gh}$ is the energy-momentum tensor and $W_{\rm gh}$ the
spin-3
current of the ghost--antighost system\foot{Later we will derive the
explicit
expressions for $T_{\rm gh}$ and $W_{\rm gh}$; $W_{\rm gh}$ will also
depend on
$T^{(1)}$ and $T^{(2)}$.}. The BRST charge is nilpotent, provided that
$c_1+c_2=100$.

To obtain a better understanding of the basic difference between the
BRST charges of [\tmieg] and [\blnw], it is instructive to restrict
oneselves to the classical $w_3$ algebra. The classical $w_3$ algebra
is obtained from the quantum $W_3$ algebra by omitting in \wthree\
all central terms and by retaining, in the OPE of the
two spin-3 generators, the quadratic terms only.
In the approach of [\tmieg]
it is assumed that
the classical $w_3$ algebra is represented by the matter sector only,
and the BRST charge corresponding to such a $w_3$ algebra is constructed.
On the other hand, in the approach of [\blnw], there is a matter and
Liouville system, corresponding to the sytems (1) and (2), respectively,
which separately satisfy the $w_3$ algebra.
We assume that $T^{(i)}$ and $W^{(1)}$ are real and that $W^{(2)}$ is
imaginary\foot{
Following the literature, we use in this paper the convention
that the matter fields are imaginary and the Liouville fields are real.
The Liouville system only satisfies the $w_3$ algebra
using an {\it imaginary}
spin-3 generator. Rescaling with a factor $i$ leads to
a real spin-3 generator but also changes the sign of the quadratic
terms in the algebra.}.
The (real) sum of the Liouville and matter sytems, i.e.
$T^{(1)}+T^{(2)}$ and $W^{(1)}+iW^{(2)}$,
does not satisfy the $w_3$ algebra,
as in the quantum case. The important point however is that, unlike
the quantum case, the sum here does satisfy a {\it modified}
$w_3$ algebra.
The difference between the two algebras resides in the quadratic terms.
The matter system satisfies a $w_3$ algebra where the quadratic
terms are proportional to
$+T^{(1)}T^{(1)}$ while the Liouville sector, using the {\it real}
generators $T^{(2)}$ and $iW^{(2)}$, satisfies
a $w_3$ algebra with an opposite sign of the quadratic terms, i.e.
they are proportional to $-T^{(2)}T^{(2)}$.
Therefore, the sum satisfies a modified $w_3$-algebra
where the quadratic terms are proportional to
$(T^{(1)})^2-(T^{(2)})^2$, i.e.
$$
\bigl (W^{(1)}+iW^{(2)}\bigr ) (z)\bigl (W^{(1)}+iW^{(2)}\bigr ) (w)
\sim
\bigl (T^{(1)}-T^{(2)}\bigr )\bigl (T^{(1)}+T^{(2)}\bigr ).\eqn\moda
$$
Since the
quadratic terms are proportional to the total spin-2 generator
$T^{(1)}+T^{(2)}$, one is left with a {\it closed} gauge algebra
which is of the so-called soft type.
In particular, we see that the combination
$T^{(1)}-T^{(2)}$ occurs as a field-dependent structure constant of the
modified algebra.
The algebra is also open in the sense that it
closes only on-shell, with the field equations given by the spin-2
and spin-3 constraints $T^{(1)}+T^{(2)}=0$ and $W^{(1)}+iW^{(2)}=0$,
respectively.

It is the aim of this letter to show how the BRST charge of [\blnw]
follows from the above-mentioned modified $w_3$ algebra. We will
first perform a classical gauging of this modified algebra and then
quantize the system. We thus derive the results of [\blnw] from a
Lagrangian point of view.

\vskip .5truecm

\noindent {\bf 2.\ Gauging}

As our starting point we consider
two sets of two scalar fields, an
imaginary one
$\varphi^{(1)}=\sum_{r=1}^2\varphi^{(1)}_rH_r$ and a real one
$\varphi^{(2)}=\sum_{r=1}^2\varphi^{(2)}_rH_r$\foot{
One can associate $\varphi^{(1)}$ with the (imaginary) matter
fields and $\varphi^{(2)}$ with the (real) Liouville fields.
In a non-critical $W_3$-string the Liouville fields describe a
$SL(3)$ Toda system.}, where
$$
H_1=\pmatrix{1&0&0\cr 0&-1&0\cr 0&0&0\cr},\qquad
H_2=\pmatrix{0&0&0\cr 0&1&0\cr 0&0&-1\cr}.\eqn\cartan
$$
The free-field action
$$
S_0=\sum_{i=1}^2\ft12
\int{\rm tr}(\pa\varphi^{(i)}{\bar\pa}\varphi^{(i)})\eqn\action
$$
transforms
under the classical $w_3$ transformations given by
$$
\eqalign{
\delta\varphi^{(1)}&=\epsilon\pa\varphi^{(1)}+
i\lambda\pa\varphi^{(1)}\varphi^{(1)}-\ft{i}3\lambda {\rm tr}
(\pa\varphi^{(1)}\pa\varphi^{(1)}), \cr
\noms
\delta\varphi^{(2)}&=\epsilon\pa\varphi^{(2)}\mp
\lambda\pa\varphi^{(2)}\varphi^{(2)}\pm \ft{1}3\lambda {\rm tr}
(\pa\varphi^{(2)}\pa\varphi^{(2)}),}\eqn\delt
$$
as
$$
\delta S_0=-\int{\bar\pa}\epsilon(T^{(1)}+T^{(2)})
-\int {\bar\pa}\lambda(W^{(1)}\pm i W^{(2)}), \eqn\deltas
$$
where
$$
\eqalign{
T^{(i)}&=-\ft12{\rm tr}(\pa\varphi^{(i)}\pa\varphi^{(i)}),\cr
W^{(i)}&=-\ft{i}3{\rm
tr}(\pa\varphi^{(i)}\pa\varphi^{(i)}\pa\varphi^{(i)})\cr}
\eqn\tandw
$$
are the spin-2 and spin-3 Noether currents. The
Poisson bracket algebra of $\{T^{(1)},W^{(1)}\}$ and
$\{T^{(2)},W^{(2)}\}$ separately is given by the $w_3$ algebra.
The Noether currents $T^{(i)}$ and $W^{(i)}$ transform as
$$
\eqalign{
\delta T^{(1)}&=\epsilon\pa T^{(1)} +2\pa\epsilon  T^{(1)}
+2\lambda\pa W^{(1)} + 3\pa\lambda W^{(1)},\cr
\noms
\delta T^{(2)}&=\epsilon\pa T^{(2)} +2\pa\epsilon  T^{(2)}
\pm 2i\lambda\pa W^{(1)} \pm 3i\pa\lambda W^{(1)},\cr
\noms
\delta W^{(1)}&=\epsilon\pa W^{(1)} +3\pa\epsilon  W^{(1)}
+\ft13\lambda\pa (T^{(1)}T^{(1)}) + \ft23\pa\lambda
(T^{(1)}T^{(1)}),\cr
\noms
\delta W^{(2)}&=\epsilon\pa W^{(2)} +3\pa\epsilon  W^{(2)}
\pm\ft{i}3\lambda\pa(T^{(2)}T^{(2)}) \pm\ft{2i}3\pa\lambda
(T^{(2)}T^{(2)}),
\cr}\eqn\deltac
$$
where we have used the fact that
$$
{\rm tr}(\pa\varphi)^4=\ft12\big({\rm
tr}(\pa\varphi)^2\big)^2.\eqn\id
$$
The factors of $i$ in the first line of \delt\  reflect the fact that
$\varphi^{(1)}$ is imaginary. The $\pm$ signs in the
second line of \delt\ are related to the invariance of the
$w_3$ algebra
under the replacement $W \rightarrow -W$ of the spin-3 generator.
This will later on explain the two possible combinations in the
BRST charge of [\blnw].

The action becomes invariant under \delt, if a minimal coupling term
$S_1$ is
added to the free action $S_0$:
$$
S_1=\int h_{(2)}(T^{(1)}+T^{(2)}) +\int  h_{(3)}(W^{(1)}\pm i
W^{(2)}),
\eqn\coupling
$$
where $h_{(2)}$ is the Beltrami differential and $h_{(3)}$ is its
spin-3 generalization\foot{
Since the Liouville fields already describe the gravity sector,
one might
wonder what the meaning is
of the additional gauge fields $h_{(2)}$ and $h_{(3)}$.
The point is that a covariant field formulation of $W_3$-gravity is
described by the spin-2 metric $g_{\mu\nu}$ and a symmetric spin-3
tensor
field $A_{\mu\nu\rho}$. The Liouville fields represent the conformal
modes of these gauge fields while $h_{(2)}$ and $h_{(3)}$
represent other components of the same gravitational gauge fields.}.
Indeed, if $h_{(2)}$ and $h_{(3)}$ transform
as
$$
\eqalign{
\delta h_{(2)} &={\bar\pa}\epsilon +\epsilon\pa h_{(2)}-\pa\epsilon
h_{(2)}
+\ft13(\lambda\pa h_{(3)}-\pa\lambda h_{(3)})(T^{(1)}-T^{(2)}),\cr
\noms
\delta h_{(3)} &={\bar\pa}\lambda +\epsilon\pa h_{(3)}-2\pa\epsilon
h_{(3)}
+2\lambda\pa h_{(2)}-\pa\lambda h_{(2)},\cr}
\eqn\deltah
$$
then $S_0+S_1$ is invariant under classical $w_3$ gauge
transformations. The
fact that minimal coupling is sufficient for gauging a chiral $w_3$
symmetry
is due to Hull [\hull]. In [\ssvannn] the chiral gauge theory was
generalized to
the case where both the chiral and antichiral $w_3$ symmetries were
gauged. It
was shown that through the introduction of auxiliary  fields, the
so-called nested
covariant derivatives, the non-chiral gauged $w_3$ symmetry reduces
to two copies
of the chiral case. It is not hard to generalize the analysis of
[\ssvannn]
to the case here at hand.

We note that the Poisson-bracket algebra of
the total currents $T^{(1)}+T^{(2)}$ and $W^{(1)}\pm iW^{(2)}$,
occurring in \coupling, is given by the modified $w_3$ algebra
mentioned in the Introduction.
The transformation rules of the gauge fields $h_{(2)}$
and $h_{(3)}$ are
determined by the structure functions of this modified algebra.
The structure of this algebra can be made explicit by
calculating the commutators of the Virasoro and spin-3 symmetries.
We find that the commutator of
two Virasoro symmetries and that of a Virasoro symmetry with a
spin-3 symmetry
still assume the standard form:
$$
\eqalign{
[\delta(\epsilon_1),\delta(\epsilon_2)]
&=\delta(\epsilon_3=\e_2\pa\e_1-\e_1\pa\e_2), \cr
\noss
[\delta(\epsilon_1),\delta(\l_2)]
&=\delta(\l_3=2\l_2\pa\e_1-\e_1\pa\l_2).\cr}\eqn\comm
$$
The difference with the usual $w_3$ algebra is only manifest
in the commutator of two spin-3 transformations. We find that this
commutator is given by
$$
\eqalign{
[\delta(\l_1),\delta(\l_2)]\varphi^{(1)}
&=\delta(\e_3)\varphi^{(1)}+\ft13(\l_2\pa\l_1-\l_1\pa\l_2){\delta
(S_0+S_1)\over \delta h_{(2)}}\pa\varphi^{(1)},\cr
\noms
[\delta(\l_1),\delta(\l_2)]\varphi^{(2)}
&=\delta(\e_3)\varphi^{(2)}-\ft13(\l_2\pa\l_1-\l_1\pa\l_2){\delta
(S_0+S_1)\over \delta h_{(2)}}\pa\varphi^{(2)},\cr
\noms
[\delta(\l_1),\delta(\l_2)]h_{(2)}
&=\delta(\e_3)h_{(2)}-\ft13(\l_2\pa\l_1-\l_1\pa\l_2)\Big\lbrace
\pa\varphi^{(1)}_i{\delta (S_0+S_1)\over \delta \varphi^{(1)}_i}
-\pa\varphi^{(2)}_i{\delta (S_0+S_1)\over \delta
\varphi^{(2)}_i}\Big\rbrace,
\cr
\noms
[\delta(\l_1),\delta(\l_2)]h_{(3)}
&=\delta(\e_3) h_{(3)},\cr}\eqn\commm
$$
where
$$
\e_3=\ft13(\l_2\pa\l_1-\l_1\pa\l_2)(T^{(1)}-T^{(2)}).\eqn\epsi
$$
The combination $T^{(1)}-T^{(2)}$, occuring in \epsi, reflects the
fact that it is a field-dependent structure
function in the modified $w_3$ algebra (cp.\ to \moda).
\vskip .5truecm

\noindent{\bf 3.\ Quantization}

To describe the quantization, it is convenient to use the
Batalin--Vilkovisky (BV) formalism [\bv]. A readable account of the BV
approach can be
found, for example in [\proeyen].
The first step in the BV formalism consists in the introduction
of extra fields, some of which are anticommuting ghost fields.
At this stage, we treat the theory still at the classical level in
the sense that all operations can be formulated in terms of
Poisson brackets.
Besides the matter fields $\varphi^{(1)}$, the Liouville fields
$\varphi^{(2)}$ and the gauge fields $h_{(2)}, h_{(3)}$, we
introduce ghost
fields $c^{(1)}$ (for the local Virasoro symmetry) and $c^{(2)}$ (for
the local spin-3
symmetry), and the corresponding antighosts $b^{(1)}$ and $b^{(2)}$,
as well as the
Nakanishi--Lautrup fields $\pi^{(1)}$ and $\pi^{(2)}$. The BV
formalism
associates with each of these fields an antifield
$\f^{(1)*},\f^{(2)*},
h^*_{(2)},h^*_{(3)},c^{(1)*},c^{(2)*},b^{(1)*},b^{(2)*},\pi^{(1)*}$
and
$\pi^{(2)*}$ of opposite statistics. The ghost number ${\rm
gh}(\Phi^*)$ of the
antifields $\Phi^*$ \foot{We denote by $\Phi^A$ all $N$ (in our
example $N=12$)
fields and by $\Phi^*_A$ all antifields.} is given by ${\rm
gh}(\Phi^*)=-1-{\rm
gh}(\Phi)$; $c^{(1)}$ and $c^{(2)}$ have ghost number $+1$ and
$b^{(1)}$ and $b^{(2)}$ ghost number $-1$, while ${\rm
gh}(\pi^{(1)})={\rm gh}(\pi^{(2)})=0$. We first extend the classical
action $S_0+S_1$ to an extended action $S[\Phi,\Phi^*]$, which is
determined by

\noindent
a)
$$
S_0+S_1=S[\Phi,0]\eqn\aa
$$
b) $S[\Phi,\Phi^*]$ satisfies the master equation:
$$
(S,S)= 2{\lap S\over\pa \Phi^A}{\rap S\over\pa \Phi^{*}_A}=0\eqn\bb
$$
c) $S$ is proper, {\it i.e.} the $2N\times 2N$ matrix
$\rap_\alpha\lap_\beta S$ has rank $N$ on-shell.

A straightforward computation yields that
$$
\eqalign{
S[\Phi,\Phi^*]=& S_0+S_1 +\int{\rm tr}\f^{(1)*}(c^{(1)}\pa\f^{(1)}
+ic^{(2)}\pa \f^{(1)}\pa \f^{(1)})\cr
&+\int{\rm tr}\f^{(2)*}(c^{(1)}\pa\f^{(2)}\mp c^{(2)}\pa \f^{(2)}\pa
\f^{(2)}) \cr
&+\int h^*_{(2)}\big({\bar\pa}c^{(1)}+c^{(1)}\pa h_{(2)}
-\pa c^{(1)} h_{(2)}+\ft13(c^{(2)} \pa h_{(3)}-\pa c^{(2)}
h_{(3)})(T^{(1)}-T^{(2)})\big) \cr
&+\int h^*_{(3)}\big({\bar\pa}c^{(2)}+c^{(1)}\pa h_{(3)}
-2\pa c^{(1)} h_{(3)}+2c^{(2)}\pa h_{(2)}-\pa c^{(2)} h_{(2)}\big)\cr
&-\int c^{(1)*}\big(c^{(1)}\pa c^{(1)}-\ft13\pa
c^{(2)}c^{(2)}(T^{(1)}-T^{(2)})\big)
-\int c^{(2)*}\big(2c^{(2)}\pa c^{(1)}+c^{(1)}\pa c^{(2)}\big)\cr
&+\int\big(b^{(1)*}\pi_{(1)}+b^{(2)*}\pi_{(2)}\big)
+\ft13\int h^*_{(2)}{\rm tr}\big(\f^{(1)*}\pa\f^{(1)}
-\f^{(2)*}\pa\f^{(2)}\big)\pa c^{(2)}c^{(2)}.\cr}\eqn\ss
$$
In order to write down the gauge-fixed action, we choose the gauge
fermion
$$
\P=\int b^{(1)}(h_{(2)}-{\hat h}_{(2)})+\int b^{(2)}(h_{(3)}-{\hat
h}_{(3)}),
\eqn\fermion
$$
where ${\hat h}_{(2)}$ and ${\hat h}_{(3)}$ are
two {\it fixed}-background Beltrami
differentials. The gauge-fixed action $S_{\rm  gf}$ is now given by
$S_{\rm gf}=S[\Phi,\Phi^*]|_{\S}$, where $\S$ is the hypersurface
determined by
$$
\Phi^*_A-{\pa\P\over\pa\Phi^A}=0.\eqn\hyper
$$
The gauge-fixed action reads
$$
\eqalign{
S_{\rm gf}=&S_0+\int(b^{(1)}{\bar\pa}c^{(1)}+b^{(2)}{\bar\pa}c^{(2)})
\cr
&+\int h_{(2)}(T^{(1)}+T^{(2)}+T_{\rm gh})
+\int h_{(3)}(W^{(1)}\pm i W^{(2)}+W_{\rm gh})\cr
&+\int\pi_{(1)}(h_{(2)}-{\hat h}_{(2)})
+\int\pi_{(2)}(h_{(3)}-{\hat h}_{(3)}),\cr}\eqn\gfaction
$$
where
$$
\eqalign{
T_{\rm gh}=&-2b^{(1)}\pa c^{(1)}-\pa b^{(1)}c^{(1)}
-3b^{(2)}\pa c^{(2)}-2\pa b^{(2)} c^{(2)},\cr
\noss
W_{\rm gh}=&-3b^{(2)}\pa c^{(1)}-\pa
b^{(2)}c^{(1)}-\ft23 b^{(1)}\pa c^{(2)}(T^{(1)}-T^{(2)})\cr
&-\ft13\pa b^{(1)} c^{(2)}(T^{(1)}-T^{(2)})-\ft13 b^{(1)} c^{(2)}\pa
(T^{(1)}-T^{(2)}).\cr}
\eqn\gcurrent
$$

By construction, $S_{\rm gf}$ is invariant under the BRST
transformations
$$
\delta\Phi^A= {\rap S[\Phi,\Phi^*]\over\pa\Phi^*_A}\Big|_{\S}\l.
\eqn\brstv
$$
Integrating over the Nakanishi--Lautrup fields $\pi_{(1)}$ and
$\pi_{(2)}$
imposes the gauge-fixing conditions $h_{(2)}={\hat h}_{(2)}$ and
$h_{(3)}=
{\hat h}_{(3)}$. The action is still BRST-invariant, but because of
the
elimination of $\pi_{(1)}$, $\pi_{(2)}$, $h_{(2)}$ and $h_{(3)}$, the
BRST transformation rules are modified by equations-of-motion terms.
The BRST
transformations are thus given by
$$
\crampest
\eqalign{
\delta\f^{(1)}&=(c^{(1)}-\ft13 b^{(1)}\pa c^{(2)}
c^{(2)})\l\pa\f^{(1)}
+ic^{(2)}\l (\pa\f^{(1)}\pa \f^{(1)}-\ft13{\rm
tr}\pa\f^{(1)}\pa\f^{(1)}),\cr
\noss
\delta\f^{(2)}&=(c^{(1)}+\ft13 b^{(1)}\pa c^{(2)}
c^{(2)})\l\pa\f^{(2)}
\mp c^{(2)}\l (\pa\f^{(2)}\pa \f^{(2)}-\ft13{\rm tr}\pa\f^{(2)}\pa
\f^{(2)}),\cr
\noss
\delta c^{(1)}&=-\big(c^{(1)}\pa c^{(1)}-\ft13\pa c^{(2)} c^{(2)}
(T^{(1)}-T^{(2)})\big)\l,\cr
\noss
\delta c^{(2)}&=-(2c^{(2)}\pa c^{(1)}+ c^{(1)}\pa c^{(2)})\l,\cr
\noss
\delta b^{(1)}&=-(T^{(1)}+T^{(2)}+T_{\rm gh})\l,\cr
\noss
\delta b^{(2)}&=-(W^{(1)}\pm iW^{(2)}+W_{\rm gh})\l,\cr}\eqn\brstvar
\uncramp
$$
and the corresponding {\it classical} BRST charge follows
immediately: it is exactly that in \brst, with $T_{\rm gh}$ and
$W_{\rm gh}$ given in
Eq. \gcurrent. It is not difficult to verify that the Poisson bracket
of two BRST
charges vanishes modulo equations-of-motion terms.
The fact that nilpotency of the BRST transformations holds only
on-shell is a well-known fact for open-gauge algebras [\twelve].

In order to promote the classical BRST charge to a charge
which also closes at the quantum level, one may proceed as follows.
One starts from the classical
action given in \gfaction\ and calculates the
possible anomalies that may arise in the quantization of this action.
They occur in two types: universal anomalies and matter-dependent
ones [\hhull]. The cancellation of these possible anomalies
leads to the addition of counter terms to the classical action.
It turns out that these counterterms occur as renormalizations of the
classical currents given in \tandw\   and \gcurrent. The quantum BRST
charge may then be obtained from the classical BRST charge, simply
by replacing every classical current by the corresponding quantum
expression.
In other words, the cancellation of all anomalies in the quantum theory
defined by the action \gfaction\ is equivalent to the
construction of a nilpotent quantum BRST operator.
This approach has been
advocated in the context of ordinary $W_3$ gravity in [\ppoo].

In this letter, we will follow another, but equivalent, approach
in which
the relevant counter terms and corresponding renormalizations of the
currents are derived from the requirement that the theory does
not depend on the
chosen gauge. We
write $S_0=S^{(1)}+S^{(2)}$, where the labels $(1)$ and $(2)$
refer to the matter and Liouville sector, respectively,
and integrate out the Nakanishi--Lautrup
fields in the classical gauge-fixed action \gfaction.
The condition that the theory does not depend on the chosen gauge
leads to the requirement
$$
{\delta W[{\hat h}_{(2)},{\hat h}_{(3)}]\over\delta {\hat h}_{(2)}}=
{\delta W[{\hat h}_{(2)},{\hat h}_{(3)}]\over\delta {\hat
h}_{(3)}}=0,\eqn\inv
$$
where the effective action $W[{\hat h}_{(2)},{\hat h}_{(3)}]$ is
defined by
$$
\crampest
e^{- W[{\hat h}_{(2)},{\hat h}_{(3)}]}=\Big\langle{\rm exp} -{1\over
\pi}\int\Big({\hat h}_{(2)}(T^{(1)}+T^{(2)}+T_{\rm gh})  +{\hat
h}_{(3)}(W^{(1)}\pm i W^{(2)}+W_{\rm gh}\Big)\Big\rangle
\eqn\eff
\uncramp
$$
To obtain the quantum expression for $T^{(1)}$ and $T^{(2)}$, we first
set ${\hat h}_{(3)}=0$ and find that
$$
W[{\hat h}_{(2)},0]={c_1+c_2-100\over 24\pi}\Gamma[{\hat
h}_{(2)}],\eqn\gam
$$
where $\Gamma[{\hat h}_{(2)}]$ is Polyakov's action for induced
gravity [\poly].
This leads to the condition that the central charges of the matter
systems should add up to 100:
$$
c_1+c_2=100.\eqn\hundred
$$
Clearly, the classical realizations of $T^{(1)}$ and $T^{(2)}$
given in Eqs. \action and \tandw\
do not satisfy this requirement. We have to add counter terms
to the classical action,  which appear as renormalizations of the
classical currents. In the case at hand, one finds that the quantum
currents that satisfy the condition \hundred\ are given by
$$
\crampest
\eqalign{
T^{(i)} =& -\ft12{\rm tr} (\pa\f^{(i)}\pa\f^{(i)})
+Q_i{\rm tr}\big(\pa^2\f^{(i)}\sum_{r=1}^2 H_r\big),\cr
\noss
W^{(i)}=& -\ft{i}3{\rm tr} (\pa\f^{(i)}\pa\f^{(i)}\pa\f^{(i)})
+iQ_i (\pa^2\f^{(i)}_1\pa\f^{(i)}_1)\cr
&-2iQ_i (\pa^2\f^{(i)}_2\pa\f^{(i)}_2) -{Q_i\over 2}\pa T^{(i)}
+ i Q_i^2\pa^3\f^{(i)}_2, \cr}\eqn\modc
\uncramp
$$
where
$$
Q_1=i\Big(\sqrt{t}+{1\over\sqrt{t}}\Big), \qquad
Q_2=\Big(\sqrt{t}-{1\over\sqrt{t}}\Big),
\quad t\in \IR.\eqn\back
$$
These quantum currents form two commuting copies of the $W_3$
algebra.

To find the appropriate renormalizations of the ghost currents,
we next consider the term in $W[{\hat h}_{(2)},{\hat
h}_{(3)}]$
with two external ${\hat h}_{(3)}$ fields and one external ${\hat
h}_{(2)}$ field. This term vanishes if and only if
$W^{(1)}\pm i W^{(2)}+W_{\rm gh}$ is
primary with
respect to $T^{(1)}+T^{(2)}+T_{\rm gh}$ at $c_1+c_2=100$.
Again, in order to achieve this, we must add local counter terms
to the action. In this case they occur as renormalizations of the
classical $W_{\rm gh}$ current
of Eq. \gcurrent. The corresponding quantum current is obtained by
adding the following additional terms to the classical current:
$$
{(17\beta_1^2-1)\over 90\beta_1^2}(2\pa^3 b^{(1)}c^{(2)}
+9\pa^2 b^{(1)}\pa c^{(2)}+15\pa  b^{(1)}\pa^2 c^{(2)}+10
b^{(1)}\pa^3 c^{(2)}).
\eqn\piece
$$
Since all counter terms occur as renormalizations of
the matter and ghost
currents, the quantum effective action is obtained from the
classical action by replacing the classical matter, Liouville and
ghost currents, by the corresponding quantum expressions which
are given in
Eqs. \modc, \gcurrent\  and \piece.
The quantum BRST charge is obtained by a similar replacement in the
classical BRST charge. One can verify that, after this
replacement, $Q$ is indeed nilpotent.
The quantum
BRST charge obtained in
this way is, up to a trivial rescaling $W^{(i)}\to W^{(i)}/\sqrt{3}$,
$W_{\rm gh}\to \pm iW_{\rm gh}/\sqrt{3}$, $h_{(3)}\to \sqrt{3}
h_{(3)}$, $c^{(2)}\to \mp i\sqrt{3} c^{(2)}$ and $b^{(2)}\to \pm
ib^{(2)}/\sqrt{3}$, precisely equal to the one introduced in [\blnw].
The sign
ambiguity in the second term of Eq.\brst\  directly follows from the
sign ambiguity in Eq.\ \delt.
The above approach of deriving the counter terms
resembles an approach that was performed
for ordinary $W_3$ gravity in [\hhull, \sev].

\vskip .5truecm

\noindent {\bf 4.\ Conclusions}

In this paper we showed how to derive the BRST charge of [\blnw] from
a Lagrangian point of view. Our basic observation is that the system of
Liouville and matter fields at the classical level is based upon a {\it
closed} gauge algebra,  which is a modification of the classical
$w_3$ algebra. Quantization  of the theory then leads exactly to the
expression of the quantum BRST charge given in [\blnw].
A novelty of this
BRST charge is that nilpotency is achieved without
the presence of a closed
quantum algebra. It would be interesting to see whether
the modified $w_3$ algebra can nevertheless be extended
to a quantum algebra in some way or another.

{\it A priori} the scalars considered in this paper
are not
necessarily designated to represent either matter or Liouville fields.
Instead, one may consider the possibility that they can all
be viewed
as matter fields so that the $W_3$ BRST charge of [\blnw] could be
used to construct
new critical $W_3$ strings. An interesting possibility
is the following.
Take a multi-scalar realization of $W_3$ [\romans]. As pointed out in
[\romans], only one of the scalars, say $\rho$, occurs
explicitly. All other scalars occur via their energy-momentum
tensor, say $T_\mu$. Furthermore, the contribution of the different
scalars to the central
charge $c_1$ is given by
$c_\rho=\ft34 c_1 - \ft12$ and $c_\mu = \ft14 c_1
+ \ft12$. Note that the critical value $c_1=100$ corresponds
to a non-critical value $c_\mu = 25\ft12$ of the Virasoro algebra.
Therefore, it seems
that one cannot embed a critical Virasoro string into a critical
$W_3$ string [\romans]. Interestingly enough, such an embedding
is possible by combining the multi-scalar realization of [\romans]
with a one-scalar realization of $W_3$ at $c_2=-2$ [\minustwo]
to form a modified
$w_3$ algebra, as described in this paper. In that case the
total central charge is given by $c_1-2$. Now the
critical value $c_1-2=100$ corresponds to the critical
value $c_\mu=26$. One thus obtains a critical $W_3$ string,
in which there
exists a critical Virasoro string with 26 free scalars.
The properties
of this critical $W_3$ string will be investigated elsewhere [\bss].

\bigskip

\ack{E.B. and A.S. thank the Theory Division at CERN for hospitality
while part
of this work was done. The work of E.B. has been made possible by
a fellowship of the Royal Netherlands Academy of Arts and
Sciences (KNAW).
The work of A.S. was supported in part by the
Director
Office of Energy Research, Office of High Energy and Nuclear Physics,
Division of High Energy Physics of the U.S. Department of Energy
under Contract
DE-AC03-765F00098, and in part by the National Science Foundation
under grant
PHY90-21139. The work of X.S. was supported
partially
by a World Laboratory scholarship. We would like to thank W. Lerche and
N.P. Warner, and especially
K. Schoutens, for useful discussions.
We performed computations using
the
meromorphic conformal field theory package OPE [\thiel].}

\refout
\end